\documentclass[aps,prl,twocolumn,floatfix,showpacs,groupaddress,longbibliography]{revtex4-2}

\usepackage{amssymb}
\usepackage{amsmath}
\usepackage{bm}
\usepackage{graphicx}
\usepackage{hyperref}
\usepackage{xcolor}
\usepackage{soul}

\usepackage[utf8]{inputenc}

\begin{document}

\title{Vortex Lattices in Active Nematics with Periodic Obstacle Arrays }

\author{Cody D. Schimming}
\email[]{cschim@lanl.gov}
\affiliation{Theoretical Division and Center for Nonlinear Studies, Los Alamos National Laboratory, Los Alamos, New Mexico, 87545, USA}

\author{C. J. O. Reichhardt}
\affiliation{Theoretical Division and Center for Nonlinear Studies, Los Alamos National Laboratory, Los Alamos, New Mexico, 87545, USA}

\author{C. Reichhardt}
\affiliation{Theoretical Division and Center for Nonlinear Studies, Los Alamos National Laboratory, Los Alamos, New Mexico, 87545, USA}

\begin{abstract}
We numerically model a two-dimensional active nematic confined by a periodic array of fixed obstacles.
Even in the passive nematic, the appearance of topological defects is unavoidable due to
planar anchoring by the obstacle surfaces.
We show that a vortex lattice state emerges as activity is increased,
and that this lattice may be tuned from ``ferromagnetic'' to ``antiferromagnetic'' by varying the gap size between obstacles.
We map the rich variety of states exhibited by the system
as a function of distance between obstacles and activity,
including
a pinned defect state, motile defects, the vortex lattice,
and active turbulence.
We
demonstrate that the flows in the active turbulent phase can 
be tuned by the presence of obstacles,
and explore the effects of a frustrated lattice geometry on the vortex lattice phase.
\end{abstract}

\maketitle
Active nematics are a  class of active fluids whose microscopic constituents are anisotropic and exert dipolar forces \cite{marchetti13,doo18}. Examples of such systems include cytoskeletal filaments with molecular motors \cite{Sanchez12}, cellular tissues \cite{saw17}, suspensions of bacteria in nematic liquid crystals \cite{zhou14b,genkin17}, and soil bacteria \cite{copenhagen21}.
One of the key features of active
nematics is that
in many cases the dynamics can
be described by the motion of defects with different topological charges
\cite{Sanchez12,DeCamp15,giomi14,Shankar19}.
There is growing interest in developing
methods to control the
defect dynamics and flows produced by active nematics,
such as by
applying external fields \cite{guillamat16},
introducing anisotropic substrates \cite{guillamat16b,thijssen20}
or imposing a
geometric confinement of the material \cite{opathalage19,thijssen21}.

In systems with well-defined length scales such
as the average distance between topological defects,
new ordered phases can arise upon coupling to a periodic
substrate \cite{Bak82}, as observed for
vortices in superconductors \cite{Harada96} and Bose-Einstein
condensates \cite{Tung06} coupled to periodic arrays,
ordering of surfaces \cite{Coppersmith82},
cold atom systems on optical substrates \cite{Bloch05}
and colloids coupled  to ordered substrates \cite{OrtizAmbriz16,Bohlein12}.
Active nematics are another system in which
commensuration effects can arise;
however, due to their nonequilibrium nature,
it should be possible for dynamical commensuration effects to
appear as well.
Here, we examine the effects of geometric confinement on active nematics induced by a periodic array of fixed obstacles for
a varied range of activity levels and obstacle sizes. 
Previous
theoretical and experimental studies of active nematic confinement
have
typically employed an external boundary as a confining structure, resulting in
a channel, circular disk, 
or annular geometry \cite{wioland13,shendruk17,wu17,norton18,opathalage19,chandragiri20,hardouin20,varghese20,schimming23b}.
As the system size and active force magnitude is varied,
such systems can exhibit
anomalous flow states such as dancing topological defects or system-wide circulation \cite{shendruk17,wu17,norton18,opathalage19}.
Relatively few studies have addressed
the effects of fixed, embedded obstacles on the active nematic flow state;
however, recent experiments have successfully produced
active nematics in obstacle laden environments \cite{figueroa22,velez23}.

Motivated by the recent experimental work, we use a minimal, active nematic continuum model to simulate an array of obstacles
that separates the system into interacting circular domains
where the obstacle shape makes the formation of topological
defects unavoidable.
We show that a variety of phases appear, including
a low active force state where the defects remain pinned to the obstacles,
a state where the defects are motile,
and an intermediate activity state where the
flow organizes into a lattice of vortices.
By varying the size of the obstacles, we can tune the vortex lattice  from ``ferromagnetic,'' in which the vortices are all of the same chirality, to ``antiferromagnetic,'' in which each nearest neighbor vortex pair is of opposite chirality. To our knowledge, this is the first report
of a ferromagnetic vortex lattice state in active nematics. We compare the active turbulent phase in systems with and without obstacles and find that the fluid flow slows as the obstacles increase in size, while the directional distribution of the fluid velocity becomes peaked along diagonal lattice directions.
Finally, we explore the effects of lattice frustration by simulating the active nematic on a triangular lattice. Our findings provide an experimentally viable method for controlling and tuning vortex lattices and flows in active nematics.

We model a two-dimensional active nematic using a well-documented continuum model that has been shown to capture the key features of experimental active nematics \cite{marenduzzo07,doo18}. In our dimensionless equations,
presented in detail in the supplemental material \cite{SuppNote23},
lengths are measured in units of the nematic correlation length $\xi$ and times are measured in units of the nematic relaxation time $\sigma$. The nematic state is captured by the tensor order parameter $\mathbf{Q} = S\left[\mathbf{n} \otimes \mathbf{n} - \left(1/2\right)\mathbf{I}\right]$ where $S$ is the scalar order parameter indicating the local degree of alignment and $\mathbf{n}$ is the director, giving the local direction of orientation. The evolution of $\mathbf{Q}$ is given by 
\begin{equation} \label{eqn:Qevo}
    \frac{\partial \mathbf{Q}}{\partial t} + \left(\mathbf{v}\cdot\nabla\right) \mathbf{Q} - \mathbf{S} = -\frac{\delta F}{\delta \mathbf{Q}}
\end{equation}
where $F$ is the Landau-de Gennes free energy with one elastic constant \cite{deGennes75}, $\mathbf{S}$ is a generalized tensor advection \cite{beris94}, and $\mathbf{v}$ is the fluid velocity. The free energy is such that the passive liquid crystal is in the nematic phase. We assume low Reynolds number flow, so the fluid velocity is given by the Stokes equation:
\begin{equation} \label{eqn:Stokes}
    \nabla^2\mathbf{v} = \nabla p + \alpha \nabla \cdot \mathbf{Q}, \qquad \nabla \cdot \mathbf{v} = 0 
\end{equation}
where  $p$ is the fluid pressure, and the last term is an addition to the usual Stokes equation that models an active force of dimensionless strength $\alpha$. The divergence free condition models incompressible flows.

\begin{figure}
    \centering
    \includegraphics[width = \columnwidth]{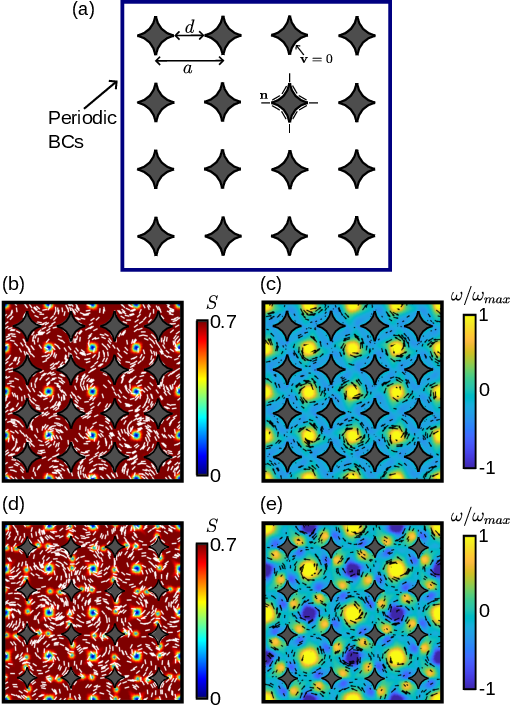}
    \caption{(a) Schematic of the computational domain with a periodic array of astroid shaped obstacles summarizing the boundary conditions on the obstacles and domain edges. (b,c) ``Ferromagnetic'' vortex lattice state
      at $d = 2$ and $\alpha = 1.8$. (b) Nematic scalar order parameter $S$ where white lines indicate the director $\mathbf{n}$. The points where $S = 0$ are topological defects. (c) Fluid vorticity with black arrows showing the velocity field.
(d) $S$ and (e) $\omega/\omega_{\rm max}$ for an ``antiferromagnetic'' vortex lattice state at $d = 6$ and $\alpha = 1.5$.}
    \label{fig:VortexLattice}
\end{figure}

Equations \eqref{eqn:Qevo} and \eqref{eqn:Stokes} are discretized in space and time and solved using the MATLAB/C++ package FELICITY \cite{walker18,SuppNote23}. We consider domains with an array of astroid shaped obstacles, as shown in Fig. \ref{fig:VortexLattice}(a). The obstacles may be thought of as separating circular domains that overlap over a distance $d$ between obstacles. In this study, we vary $d$ from $1$ to $10$
while fixing the distance between centers of obstacles to $a = 14$ so that the overall system size remains constant. The limit $d = 0$ gives individual, non-interacting circular domains, while $d = a = 14$ indicates a system with no obstacles. We employ periodic boundary conditions on the side edges of the domain, while for the obstacles we impose strong planar anchoring for $\mathbf{Q}$ and no-slip conditions for $\mathbf{v}$. The shape of the inclusions and the strong planar anchoring necessitate the formation of topological defects, or points where the nematic director is singular.
Specifically, there must be a total topological charge (winding number) of $+N$ for $N$ obstacles because each obstacle carries a charge of $-1$ that must be accounted for in the nematic. Therefore, topological defects inherently exist in the system even in the passive state with $\alpha = 0$.  

Upon increasing $\alpha$ for various obstacle sizes,
we typically find three qualitative transitions. At zero and small activity, the topological defects are pinned to the obstacles. As the activity is increased, the defects begin to unpin and move from obstacle to obstacle with little to no unbinding of new defects. When activity is further increased, a central vortex forms in each circular domain with two $+1/2$ topological charge defects encircling one another, as has been shown previously for individual circular domains \cite{norton18,opathalage19,schimming23b}. At higher activities the $+1/2$ defects merge and a stable $+1$ spiral defect forms in the domain center. Figure \ref{fig:VortexLattice}(b--e) shows examples of the nematic order parameter and velocity field in the vortex lattice state, which can have either ferromagnetic or antiferromagnetic order. At still higher activities the central vortex in each domain is no longer stable and an active turbulent phase persists in which defects are constantly unbinding and annihilating (see Supplemental Movie 1). 

\begin{figure}
    \centering
    \includegraphics[width = \columnwidth]{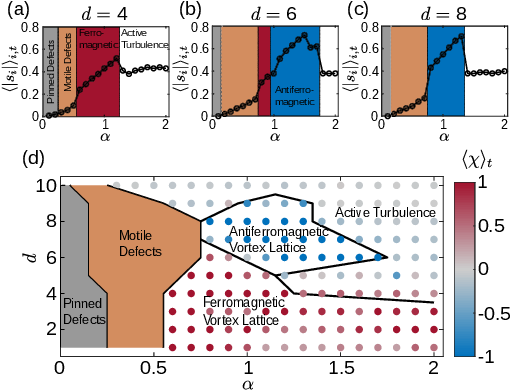}
    \caption{(a--c) Average vortex ``spin'' $\langle|s_i|\rangle_{i,t}$ as a function of activity $\alpha$ for obstacle gaps (a) $d = 4$, (b) $d = 6$, and (c) $d = 8$. Colors indicate the phase identity. (d) Phase diagram as a function of $d$ vs $\alpha$, where dot color indicates the value
of
the time-averaged spin-spin correlation function $\langle \chi\rangle_{t}$. $\langle \chi\rangle_t$ is not well-defined in the pinned defect or motile defect phases.}
    \label{fig:PhaseD}
\end{figure}

To better quantify the transitions discussed above, we map our system to a lattice of vortex ``spins'' by measuring the average vorticity in a circle of diameter $a/2$ at the center of each circular domain. The spins are then indexed by their lattice position $s_{i}$. Figure \ref{fig:PhaseD}(a--c) shows plots of $\langle|s_i|\rangle_{i,t}$, where $\langle \cdot \rangle_{i,t}$ denotes an average over lattice sites and time, as a function of activity for obstacle gaps $d=4$, $d=6$, and $d=8$. The transition to the vortex lattice state is marked by a jump followed by a linear increase in $\langle|s_i|\rangle_{i,t}$. A second transition is marked by an abrupt decrease in $\langle|s_i|\rangle_{i,t}$ which remains roughly constant. This is the active turbulent phase. 

We first focus on the vortex lattice phase. As shown in Fig. \ref{fig:VortexLattice}(b--e), vortex lattices appear with either ``ferromagnetic'' or ``antiferromagnetic'' order. To quantify this order we measure the spin-spin correlation function
\begin{equation}
    \chi = \frac{\sum_{\left\langle i,j\right\rangle} s_i s_j}{\sum_{\left\langle i,j\right\rangle} |s_i s_j|}
\end{equation}
where $\sum_{\left\langle i,j\right\rangle}$ denotes a sum over nearest neighbor pairs. For perfectly ferromagnetic order $\chi = 1$ while for perfectly antiferromagnetic order $\chi = -1$. In Fig.~\ref{fig:PhaseD}(d) we plot the boundaries of the phases as well as $\langle \chi\rangle_{t}$ as a function of activity and obstacle gap for systems in both the vortex lattice and active turbulence regimes. We do not define $\chi$ for small activities since the central vortices, and hence spins, are not well-established in the pinned defect and motile defect phases. As the obstacle gap increases, there is a window of activity values where the vortex lattice abruptly transitions from ferromagnetic to antiferromagnetic. In this window, the obstacle size can be used to tune the state of the vortex lattice. For large obstacle gaps the vortex lattice phase disappears and only active turbulence occurs.
At the transition
from vortex lattice to active turbulence,
$\langle \chi \rangle_t$ decreases in size but typically
maintains the same sign, indicating that
vestigial ferromagnetic or antiferromagnetic order
persists until $\alpha$ becomes large enough that
$\langle \chi \rangle_{t} \to 0$.

In previous work on active nematics, a one-dimensional antiferromagnetic vortex lattice was observed in channel confinement geometries \cite{shendruk17}, while numerical predictions indicate that a two-dimensional antiferromagnetic vortex lattice should appear in systems with large enough substrate friction \cite{doo16}. In our system, we assume zero substrate friction so the vortices are stabilized purely by geometric confinement, but we expect the vortex lattice phase to be stable against inclusion of some substrate friction based on recent work on circularly confined systems \cite{schimming23b}.
We find both ferromagnetic and antiferromagnetic vortex lattices
depending on the values of $d$ and $\alpha$,
and to our knowledge, a ferromagnetic vortex lattice has not
previously been observed or predicted in active nematics.

Vortex lattices can also form in bacterial suspensions confined by pillar arrays, but in that system, hydrodynamic interactions are the dominant ordering mechanism, and therefore the vortex lattice typically has antiferromagnetic ordering \cite{wioland16,reinken20}. In contrast, in the active nematic considered here, elastic forces play the dominant role in ordering. As shown in Fig. \ref{fig:VortexLattice}(b), since the central $+1$ defect that forms is of spiral type, the elastic energy between circular domains is minimized if each domain has the same chirality. Thus, elastic interactions promote ferromagnetic order.

The transitions to antiferromagnetic order and active turbulence may be explained by a competition between elastic and active forces and a hierarchy of length scales. The active nematic length scale $\xi_a \propto 1/\sqrt{\alpha}$ sets the defect density of a bulk active nematic system \cite{doo18}. For a given obstacle size, there is an effective length scale associated with the circular domains $R_{\rm eff}$. If $\xi_a \lesssim R_{\rm eff}$,
there is enough space to nucleate defects and
reach the optimal defect density, so
the system transitions to an active turbulent state.
There is, however, another length scale associated with the size of the obstacles: the obstacle gap $d$. In Fig. \ref{fig:VortexLattice}(d) the antiferromagnetic vortex lattice contains extra $\pm 1/2$ defect pairs that sit in the obstacle gaps and mediate the change in chirality between central vortices. If $\xi_a \gtrsim d$, a defect pair will not be stable in the gap. On the other hand, if $d \gtrsim \xi_a \gtrsim R_{\rm eff}$, the vortex lattice phase is stable since an extra defect pair can nucleate in the gaps to mediate the antiferromagnetic vortex order. We note that if $d < R_{\rm eff}$, this hierarchy of length scales cannot occur, explaining why we do not observe antiferromagnetic vortex order for small $d$.

We now turn our attention to the active turbulent phase that occurs for large $\alpha$. As mentioned above, the active turbulent regime is associated with a sharp decrease in $\langle |s_i| \rangle_{i,t}$ (Fig. \ref{fig:PhaseD}). We can also measure the average number of defects $\langle N_D \rangle_{t}$ to detect the transition (see the Supplemental Material for details on how this and other measures are computed \cite{blow14,schimming22,SuppNote23}). In the vortex lattice phase, $\langle N_D \rangle_{t}$ is roughly constant, while in active turbulence $\langle N_D \rangle_{t}$ grows linearly with $\alpha$ (Fig. S1). Both measures are consistent with one another in marking the transition and we use them to determine the boundary shown in Fig.~\ref{fig:PhaseD} between the motile defect phase and active turbulence for large $d$.

\begin{figure}
    \centering
    \includegraphics[width = \columnwidth]{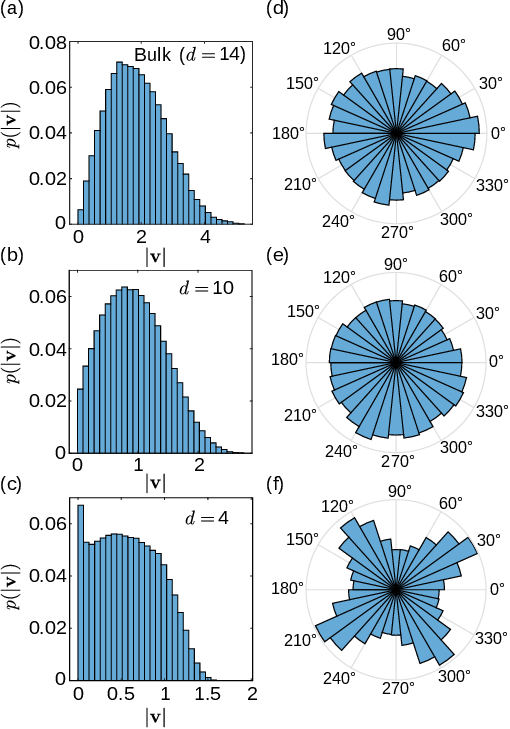}
    \caption{Flow velocity distributions in the active turbulent phase with $\alpha = 1.5$ and various obstacle gap sizes. (a--c) Distributions of $|\mathbf{v}|$ for (a) a bulk system with no obstacles, (b) obstacle gap $d = 10$, and (c) obstacle gap $d = 4$. (d--e) Velocity direction distributions $p(\theta_v)$ for (d) a bulk system with no obstacles, (e) obstacle gap $d = 10$, and (f) obstacle gap $d = 4$.}
    \label{fig:VelocityDistributions}
\end{figure}

It is instructive to compare the active turbulent phase of the obstacle array system with the $d=a$ bulk system free of obstacles.
While the flows in all systems become decorrelated over long time scales (Fig. S2), the flow velocity distributions $p(|{\bf v}|)$ vary. In Fig.~\ref{fig:VelocityDistributions}(a--c) we
plot $p(|\mathbf{v}|)$ for
systems with $\alpha = 1.5$ at $d=14$ (the bulk system), $d=10$, and $d=4$.
Figure~\ref{fig:VelocityDistributions}(a) indicates that
$p(|{\bf v}|)$ has
a two-dimensional Maxwell-Boltzmann distribution
with a maximum weight that shifts toward $|{\bf v}|=0$ as
the obstacle size increases and $d$ becomes
smaller.
This is a natural consequence of the fact that the obstacles
  have no-slip conditions and their surface area
increases as $d$ decreases.
The corresponding velocity direction distributions $p(\theta_v)$
in Fig.~\ref{fig:VelocityDistributions}(d--f),
where $\theta_v=\tan^{-1}({v_y/v_x})$,
show an isotropic distribution for small obstacles ($d=10$)
that is nearly identical to
$p(\theta_v)$ for the bulk system.
For larger obstacles,
$p(\theta_v)$ becomes anisotropic and peaks along the lattice diagonals. These velocity statistics suggest that immersed obstacles can provide
control over the flows even in the active turbulent phase,
which could contribute to the development of novel microfluidic
devices composed of active fluids.

\begin{figure}
    \centering
    \includegraphics[width = \columnwidth]{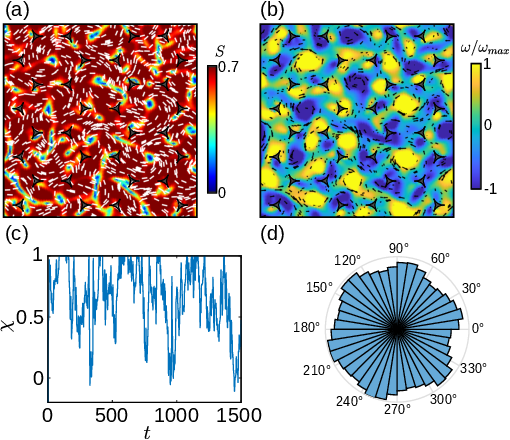}
    \caption{Active nematic system in a honeycomb lattice of concave triangular obstacles. (a) Time snapshot of the nematic configuration at $a = 8$, $d = 4$, and $\alpha = 2$ with color given by the scalar order parameter $S$ and white lines indicating the nematic director $\mathbf{n}$. (b) The corresponding vorticity (color) and flow velocity (black arrows).
(c) Spin-spin correlation function $\chi$ vs time $t$.
(d) Velocity direction distribution $p(\theta_v)$.}
    \label{fig:TriLattice}
\end{figure}

Finally, we simulate the active nematic in a domain constrained by
a honeycomb lattice of concave, triangular obstacles similar to those used in recent experiments \cite{figueroa22}.
Here, the obstacles introduce a triangular lattice of circular domains. An example nematic configuration and flow profile is shown in Fig. \ref{fig:TriLattice}(a,b).
In order to promote antiferromagnetic ordering, we
place the system just outside the ferromagnetic vortex lattice phase for this geometry: $a = 8$, $d = 4$, and $\alpha = 2$;
however, the frustration in the lattice
prevents the emergence of antiferromagnetic vortex order.
Instead we find
a primarily ferromagnetic state in which the competition between
elastic and active forces results in constantly flipping spins and the formation
of
a dynamical state similar to active turbulence. In Fig. \ref{fig:TriLattice}(c) we plot $\chi(t)$ over the course of a simulation.
There are multiple time intervals where
$\chi = 1$, indicating a perfectly ferromagnetic vortex lattice;
however, spin flips generated by the unbinding of new defects constantly reduce $\chi$, which sometimes becomes negative.
While the dynamics resemble
active turbulence, we argue that they are actually
closer to those of the vortex lattice state in the square lattice with $d=6$ and $\alpha = 0.9$, in which the underlying vortex lattice order
is ferromagnetic but defect unbinding
pushes the system towards antiferromagnetic order. We show in Fig. S3 that the $\chi$-$\chi$ temporal autocorrelation function for the frustrated triangular lattice is similar to that of the square lattice with $d=6$ and $\alpha = 0.9$. In both systems, but unlike an active turbulent system, $\chi$ is correlated over long times, indicating that spins and spin flips are also correlated over time.
In Fig.~\ref{fig:TriLattice}(d) we show that $p(\theta_v)$
peaks at the six diagonals of the lattice.
We expect the peaks to become more prominent as the obstacle size
increases.

{\it Conclusion---}
We have numerically studied the effects
on active nematics
of fixed periodic astroid shaped
obstacles with planar nematic anchoring.
As a function of activity and obstacle size,
we find a wide variety of phases,
including a pinned defect phase, motile defect regime, and
a vortex lattice phase that
can be tuned from
ferromagnetic to antiferromagnetic.
There is an active turbulent phase
that displays unique anisotropic velocity distributions at high activities, suggesting a new method to
control active turbulence with obstacles. We also find that an antiferromagnetic vortex lattice on a triangular lattice exhibits an active frustrated state. Our system should be experimentally realizable using existing approaches for creating obstacles in active nematic  systems \cite{figueroa22,velez23}.
Future directions are to consider obstacle shapes that stabilize other kinds of topological defects. Also, different lattices may yield even more exotic flow states, which opens the prospect of flow control in active nematics using obstacles.

\begin{acknowledgements}
This work was supported by the U.S. Department of Energy through the Los Alamos National Laboratory. Los Alamos National Laboratory is operated by Triad National Security, LLC, for the National Nuclear Security Administration of the U.S. Department of Energy (Contract No. 89233218CNA000001).
\end{acknowledgements}

\bibliography{LC}

\end{document}


\title{Supplemental Material for ``Vortex Lattices in Active Nematics with Periodic Obstacle Arrays''}

\author{Cody D. Schimming}
\email[]{cschim@lanl.gov}
\affiliation{Theoretical Division and Center for Nonlinear Studies, Los Alamos National Laboratory, Los Alamos, New Mexico, 87545, USA}

\author{C. J. O. Reichhardt}
\affiliation{Theoretical Division and Center for Nonlinear Studies, Los Alamos National Laboratory, Los Alamos, New Mexico, 87545, USA}

\author{C. Reichhardt}
\affiliation{Theoretical Division and Center for Nonlinear Studies, Los Alamos National Laboratory, Los Alamos, New Mexico, 87545, USA}

\maketitle

\section{Continuum active nematic model}
We numerically solve a set of partial differential equations that model the time evolution of the nematic tensor order parameter $\mathbf{Q}$ and the fluid velocity $\mathbf{v}$. The tensor order parameter may be parameterized as $\mathbf{Q} = S\left[\mathbf{n}\otimes\mathbf{n} - (1/2)\mathbf{I}\right]$ where $S$ is the degree of alignment of the nematogens and $\mathbf{n}$ is the nematic director. The equations are given by \cite{marenduzzo07,doo18}:
\begin{gather}
\frac{\partial \mathbf{Q}}{\partial t} + \left( \mathbf{v} \cdot \nabla\right) \mathbf{Q} - \mathbf{S} = -\frac{1}{\gamma} \frac{\delta F}{\delta \mathbf{Q}} \label{eqn:Qevo} \\
\eta \nabla^2 \mathbf{v} = \nabla p + \alpha \nabla \cdot \mathbf{Q} \label{eqn:Stokes}\\
\nabla \cdot \mathbf{v} = 0 \label{eqn:Incompressible}
\end{gather}
where $F$ is the free energy of the nematic, $\gamma$ is a rotational viscosity, $\eta$ is the fluid viscosity, $\alpha$ is the strength of the active force, and 
\begin{equation}
    \mathbf{S} = \left(\lambda\mathbf{E} + \bm{\Omega}\right)\left(\mathbf{Q} + \frac{1}{2}\mathbf{I}\right) + \left(\mathbf{Q} + \frac{1}{2}\mathbf{I}\right)\left(\lambda\mathbf{E} - \bm{\Omega}\right) - 2\lambda\left(\mathbf{Q} + \frac{1}{2}\mathbf{I}\right)\left(\nabla\mathbf{v} : \mathbf{Q}\right)
\end{equation}
is a generalized tensor advection\cite{beris94}. $\mathbf{E} = \left(\nabla \mathbf{v} + \nabla \mathbf{v}^T\right)/2$ is the strain rate tensor, $\bm{\Omega} = \left(\nabla \mathbf{v} - \nabla \mathbf{v}^T\right)/2$ is the rotation tensor, and $\lambda$ is a dimensionless parameter that determines whether the liquid crystal is flow aligning or flow tumbling and depends on the geometry of the nematogen. $\alpha > 0$ models an extensile active force while $\alpha < 0$ models a contractile active force. The free energy we use is the two-dimensional Landau-de Gennes free energy:
\begin{equation}
    F = \int \left(A |\mathbf{Q}|^2 + C |\mathbf{Q}|^4 + L |\nabla \mathbf{Q}|^2\right) \, d\mathbf{r}
\end{equation}
where $A$ and $C > 0$ are phenomenological material parameters and $L$ is an elastic constant.

We non-dimensionalize the system by working in dimensionless length and time,
\begin{equation}
    \mathbf{\tilde{r}} = \frac{\mathbf{r}}{\xi}, \quad \tilde{t} = \frac{t}{\sigma}, \quad \xi = \sqrt{\frac{\varepsilon L}{C}}, \quad \sigma = \frac{\gamma}{C \xi^2}
\end{equation}
where $\xi$ is the nematic correlation length, $\sigma$ is the nematic relaxation time, and $\varepsilon$ is a dimensionless parameter that controls the size of defects. The free energy is then written in units of $C\xi^2$ and Eq. \eqref{eqn:Stokes} may be multiplied by $\xi / C$ to be put in a non-dimensional form. This yields the dimensionless quantities
\begin{equation}
    \tilde{F} = \frac{F}{C\xi^2}, \quad \tilde{A} = \frac{A}{C}, \quad \tilde{p} = \frac{p}{C}, \quad \tilde{\eta} = \frac{\eta}{\sigma C}, \quad \tilde{\alpha} = \frac{\alpha}{C}.
\end{equation}

There are three important length scales associated with the above model that interact with the confining length scales of the domain. The first is the nematic correlation length $\xi$, which is the length over which nematic distortions occur. All other lengths in our system are compared with $\xi$. The size of defects, which is set by the parameter $\varepsilon$, is also an important length scale which we fix in simulations. Finally, the active length, $\xi_a = \sqrt{L/\alpha}$ is the length at which active forces perturb the underlying nematic. It is correlated with the average distance between defects and inversely related to defect density. In dimensionless units it is given by $\tilde{\xi_a} = \sqrt{1/\varepsilon \tilde{\alpha}}$. In the main text, and in following sections, the tildes denoting dimensionless quantities are omitted for brevity.

\section{Numerical Implementation}
To numerically solve Eqs. \eqref{eqn:Qevo}, \eqref{eqn:Stokes}, and \eqref{eqn:Incompressible}, we first write $\mathbf{Q}$ in a basis for two dimensional traceless, symmetric tensors
\begin{equation}
    \mathbf{Q} = \begin{pmatrix}
                 q_1 & q_2 \\
                 q_2 & -q_1
                 \end{pmatrix}
\end{equation}
and rewrite Eq. \eqref{eqn:Qevo} in terms of $q_1$ and $q_2$. Eqs. \eqref{eqn:Qevo}, \eqref{eqn:Stokes}, and \eqref{eqn:Incompressible} are then written in weak form, where the pressure is used as a Lagrange multiplier to impose the incompressibility condition.

We use the Matlab/C++ package FELICITY \cite{walker18} to generate the finite element matrices from the weak form of Eqs. \eqref{eqn:Qevo}, \eqref{eqn:Stokes}, and \eqref{eqn:Incompressible}. The time evolution of $\mathbf{Q}$ is discretized using a backward-Euler method with time step $\delta t = 0.5$. We set $A = -0.5$ so the passive liquid crystal is in the nematic phase with $S_N = \sqrt{1/2}$, the scalar order parameter that minimizes the free energy, and $\varepsilon = 4$ so defects have diameter roughly unity. We also set $\lambda = 1$ so the nematic is flow aligning. The outer boundaries of the domain are given periodic conditions while the boundaries of the fixed obstacles in the domain are set such that $S = S_N$, $\mathbf{n} = \mathbf{T}_0$, where $\mathbf{T}_0$ is the local tangent vector to the boundary curve, and $\mathbf{v} = 0$. The system is initialized with a random director field except at the fixed boundaries. At each time step, Eq. \eqref{eqn:Stokes} is first solved to give the velocity field for the given nematic configuration. The discretized, nonlinear equations for $\mathbf{Q}$, Eq. \eqref{eqn:Qevo}, are then solved using a Newton-Raphson method. We simulate in the range of $500$ to $3000$ time steps to allow each system to reach either a stationary or dynamical steady state. We define a dynamical steady state as a state in which the time average of measured quantities does not appreciably change if taken over later times.

\section{Measures}

There are several measures used for the analysis of the phases described in the main text. Here we explicitly define all measures used and give some context for their interpretation.

The primary measure used to analyze the system is the vortex ``spin.'' At each time step, the vortex spins are measured by taking the average vorticity in a circle of radius $a/4$ at the center of each circular domain:
\begin{equation}
    s_i(t) = \frac{16}{\pi a^2} \int_{A_i} \omega(\mathbf{r},t) \, d\mathbf{r}
\end{equation}
where $a$ is the distance between obstacle centers and $A_i$ is the circle of radius $a/2$ centered on the $i$th circular ``lattice site'' (see Fig. 1 in the main text). The average $\langle |s_i|\rangle_{i,t}$ may then be used as a succinct measure to demarcate the boundaries between the motile defect, vortex lattice, and active turbulence phases, as shown in Fig. 2 in the main text. We note that the computed time averages are only over times in which the system is in a stationary or dynamical steady state. 

To further quantify the system in the vortex lattice and active turbulence phases we compute the spin-spin correlation function
\begin{equation}
    \chi(t) = \frac{\sum_{\langle i,j\rangle} s_i(t) s_j(t)}{\sum_{\langle i,j\rangle} |s_i(t) s_j(t)|}
\end{equation}
where the sum is only over nearest neighbor lattice sites. The spin-spin correlation function gives the degree of spatial correlation of the vortex lattice, with $\chi = +1$ a perfectly ferromagnetic state and $\chi = -1$ a perfectly antiferromagnetic state. For each simulation we compute the time average $\langle \chi \rangle_t$ to determine the degree of order in the vortex lattice state.

The number of topological defects in the system is also a useful metric, particularly when determining the onset of active turbulence. To compute the number of topological defects, we first compute the topological defect density from $\mathbf{Q}$ \cite{blow14,schimming22},
\begin{equation}
    D(\mathbf{r},t) = \varepsilon_{i j}\varepsilon_{k \ell} \partial_k Q_{i n}(\mathbf{r},t) \partial_{\ell} Q_{j n}(\mathbf{r},t)
\end{equation}
where summation over repeated indices is assumed and $\bm{\varepsilon}$ is the two-dimensional antisymmetric tensor. The number of defects is then computed as
\begin{equation}
    N_D(t) = \frac{1}{\pi S_N^2} \int_{\Omega} |D(\mathbf{r},t)| \, d\mathbf{r}
\end{equation}
where the integration is taken over the whole domain. Note that this definition counts half integer defects as a single defect, so integer defects are counted as two. In the ferromagnetic phase, each circular domain has a $+1$ spiral defect in the center so $N_D = 32$ for $16$ obstacles. On the other hand, in the antiferromagnetic phase, in addition to the $+1$ defects at the center of each domain there is a $\pm 1/2$ defect pair in between each domain, leading to $N_D = 96$ for $16$ obstacles. In Fig. \ref{fig:NumDefects} we show the time average $\langle N_D \rangle_t$ as a function of activity $\alpha$ for various obstacle gaps. $\langle N_D \rangle_t$ increases linearly with $\alpha$ in the active turbulence phase and thus may be used as an effective measure to locate the active turbulence phase.

We compare the dynamics of the active turbulence phase for various obstacle gaps by measuring the velocity autocorrelation function
\begin{equation}
    C_{\mathbf{v}\mathbf{v}}(\tau) = \left\langle\frac{\mathbf{v}(\mathbf{r},t + \tau) \cdot \mathbf{v}(\mathbf{r},t)}{|\mathbf{v}(\mathbf{r},t)|^2}\right\rangle_{\mathbf{r},t}.
\end{equation}
Figure \ref{fig:VVcorrelations} shows $C_{\mathbf{v}\mathbf{v}}(\tau)$ for a bulk system with no obstacles, a system with obstacle gap $d = 10$, and a system with obstacle gap $d=4$. All of the systems shown are in the active turbulence phase and all three have similar velocity autocorrelation functions. For smaller obstacle gap (i.e. larger obstacles) the system is slightly more correlated at long times because of the tendencies for flows to be trapped in the domains for longer, hence there is less system-wide mixing.

Finally, to compare the dynamics of the triangular lattice and square lattice and to compare the dynamics of the vortex lattice phase and active turbulence phase, we compute the temporal autocorrelation function of $\chi$:
\begin{equation}
    C_{\chi \chi}(\tau) = \left\langle\frac{\chi(t + \tau) \chi(t)}{\chi^2(t)}\right\rangle_t.
\end{equation}
In the vortex lattice phase $\chi$ is correlated in time, even if it is fluctuating. On the other hand, $\chi$ quickly becomes decorrelated in the active turbulence phase. In Fig. \ref{fig:ChiChiCorrelations} we show $C_{\chi\chi}$ for the case of a system with a triangular domain lattice (with $a = 8$, $d = 4$, and $\alpha = 2$), a system with a square domain lattice (with $a = 14$, $d = 6$, and $\alpha = 0.9$), and the same square lattice system at higher activity, $\alpha = 2$. The square lattice system at $\alpha = 0.9$ is in the vortex lattice phase while at $\alpha = 2$ the system is in the active turbulence phase. We note that the triangular lattice phase remains correlated at long times and is similar to the case of the square vortex lattice phase, even though the qualitative dynamics are similar to active turbulence. This indicates that the spin flips in the system are correlated in time and that the frustration in the lattice promotes the dynamics. 

\section{Supplemental Movies}
\begin{itemize}
    \item Supplemental Movie 1: Active turbulent dynamics in a system with a square lattice of obstacles at $a = 14$, $d = 4$, and $\alpha = 2$. (Left) Nematic scalar order parameter $S$ with white lines giving director $\mathbf{n}$. (Right) Vorticity $\omega$ with black arrows indicating velocity direction and magnitude.
    \item Supplemental Movie 2: Dynamics of a system with a honeycomb lattice of obstacles that gives rise to a triangular lattice of circular domains at $a = 8$, $d = 4$, and $\alpha = 2$. (Left) Nematic scalar order parameter $S$ with white lines giving director $\mathbf{n}$. (Right) Vorticity $\omega$ with black arrows indicating velocity direction and magnitude.
\end{itemize}

\section{Supplemental Figures}

\renewcommand{\thefigure}{S1}
\begin{figure}[h]
    \centering
    \includegraphics[width = \textwidth]{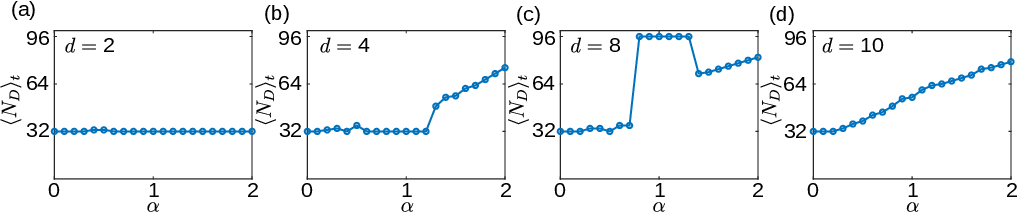}
    \caption{Average number of defects $\langle N_D\rangle_t$ versus activity $\alpha$ for obstacle gaps (a) $d = 2$, (b) $d = 4$, (c) $d=8$, and (d) $d = 10$. $N_D = 32$ in the ferromagnetic vortex lattice phase while $N_D = 96$ in the antiferromagnetic vortex lattice phase. $\langle N_D\rangle_t$ grows linearly with $\alpha$ in the active turbulence phase.}
    \label{fig:NumDefects}
\end{figure}

\renewcommand{\thefigure}{S2}
\begin{figure}[h]
    \centering
    \includegraphics[width = \textwidth]{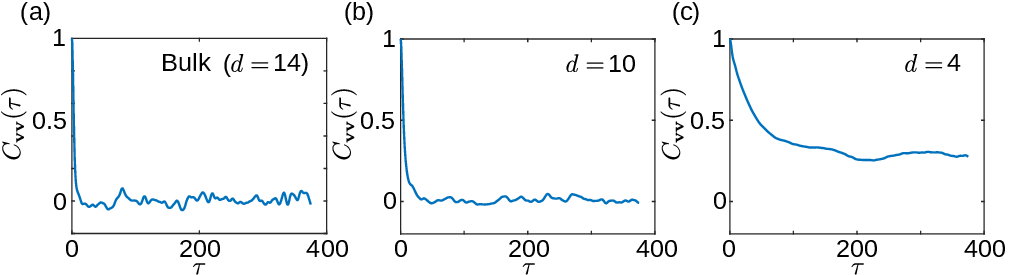}
    \caption{Velocity autocorrelation function $C_{\mathbf{v}\mathbf{v}}(\tau)$ versus advanced time $\tau$ for systems with (a) no obstacles, (b) obstacle gap $d = 10$, and (c) obstacle gap $d = 4$. For all simulations shown $\alpha = 2$.}
    \label{fig:VVcorrelations}
\end{figure}

\renewcommand{\thefigure}{S3}
\begin{figure}[h]
    \centering
    \includegraphics[width = \textwidth]{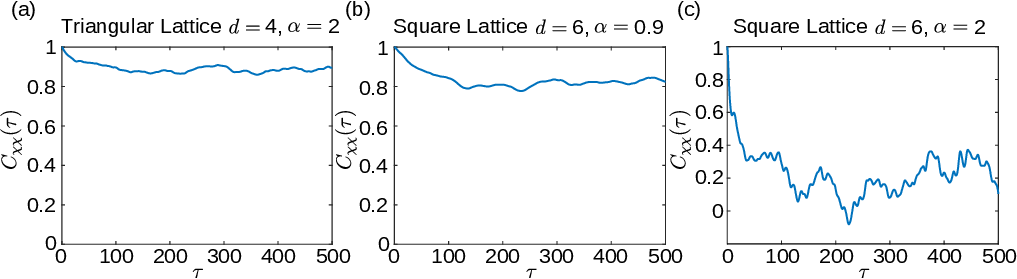}
    \caption{Temporal autocorrelation function of $\chi$, $C_{\chi\chi}(\tau)$ versus advanced time $\tau$ for (a) a triangular lattice system with $a=8$, $d=4$, and $\alpha = 2$, (b) a square lattice system with $a=14$, $d=6$, and $\alpha = 0.9$, and (c) a square lattice system with $a=14$, $d=6$, and $\alpha = 2$.}
    \label{fig:ChiChiCorrelations}
\end{figure}

\newpage

\bibliography{LC}